\documentclass[10pt]{article}
\usepackage{geometry}
\usepackage{xcolor}
\definecolor{graycolor}{gray}{0.9} 
\usepackage{microtype}
\usepackage{setspace} 
\usepackage[utf8]{inputenc}
\usepackage[english]{babel}
\usepackage{times}
\usepackage{array}
\usepackage{soul}
\usepackage{cite}
\usepackage[numbers,sort&compress]{natbib}
\usepackage{natbib}

\usepackage{titlesec} 
\titleformat {\section} [block] {\raggedright \fontsize{10}{10}\selectfont\bfseries} {\thesection. \space} {0pt} {}
\titlespacing {\section} {0pt} {12pt} {6pt}
\titleformat {\subsection} [block] {\raggedright \fontsize{10}{10}\selectfont\itshape} {\thesubsection .\space} {0pt} {}
\titlespacing {\subsection} {0pt} {12pt} {6pt}
\titleformat {\subsubsection} [block] {\raggedright \fontsize{10}{10}\selectfont} {\thesubsubsection .\space} {0pt} {}
\titlespacing {\subsubsection} {0pt} {12pt} {6pt}
\titleformat {\paragraph} [block] {\raggedright \fontsize{10}{10}\selectfont} {} {0pt} {}
\titlespacing {\paragraph} {0pt} {12pt} {6pt}

\usepackage{array} \newcommand{\PreserveBackslash}[1]{\let\temp=\\#1\let\\=\temp}
\newcolumntype{C}[1]{>{\PreserveBackslash\centering}m{#1}}
\newcolumntype{R}[1]{>{\PreserveBackslash\raggedleft}m{#1}}
\newcolumntype{L}[1]{>{\PreserveBackslash\raggedright}m{#1}}
\usepackage{lineno}
\usepackage{tabularx}
\usepackage{colortbl}
\usepackage{graphicx}
\usepackage{float}
\usepackage[export]{adjustbox}
\usepackage{caption}
\usepackage{fancyhdr} 
\setcitestyle{open={[},close={]},citesep={,\!},numbers}
\usepackage{amsmath,amssymb,amsthm,bm}
\usepackage{authblk}

\newcommand{\dd}{\mathrm{d}}
\newcommand{\e}{\mathrm{e}}
\newcommand{\eff}{\mathrm{eff}}
\newcommand{\BH}{\mathrm{BH}}
\newcommand{\bhor}{\mathrm{b}}
\newcommand{\chor}{\mathrm{c}}
\newcommand{\loc}{\mathrm{loc}}

\usepackage{lastpage}
\usepackage{layout}
\usepackage{setspace} 
\usepackage{enumitem}
\usepackage{booktabs}
\usepackage{arydshln}
\usepackage{multirow}
\usepackage{color}
\usepackage{hyperref} 
\hypersetup{
	colorlinks=true,
	linkcolor=blue,
	filecolor=blue,
	urlcolor=black,
	citecolor=cyan,
}

\fancypagestyle{firstpage}{
    \setlength{\headsep}{2.2cm}
    
    \setlength{\footskip}{1.5cm}
    \fancyhf{}
    \lhead{\begin{table}[H]
        \centering
        \begin{tabular}{L{2.5cm}C{10cm}C{3.1cm}R{2cm}}
            \includegraphics[scale=0.035]{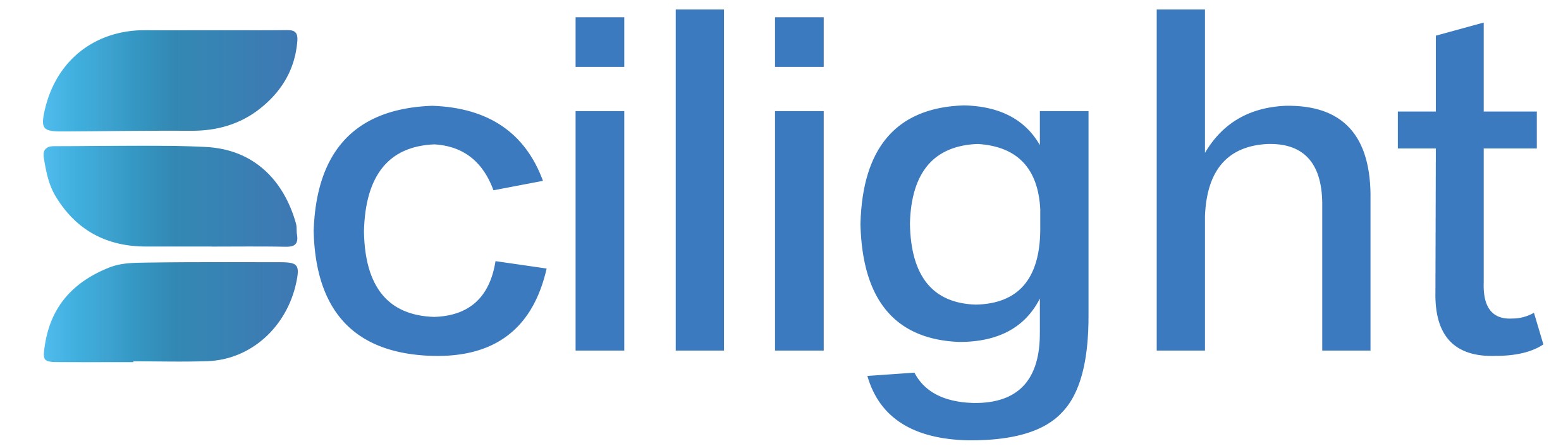} \vspace{-6pt}& \cellcolor{graycolor}\begin{tabular}[c]{@{}c@{}}\textit{International Journal of Gravitation and Theoretical Physics}\\ \href{https://www.sciltp.com/journals/ijgtp}{https://www.sciltp.com/journals/ijgtp}\end{tabular} & \includegraphics[scale=0.012]{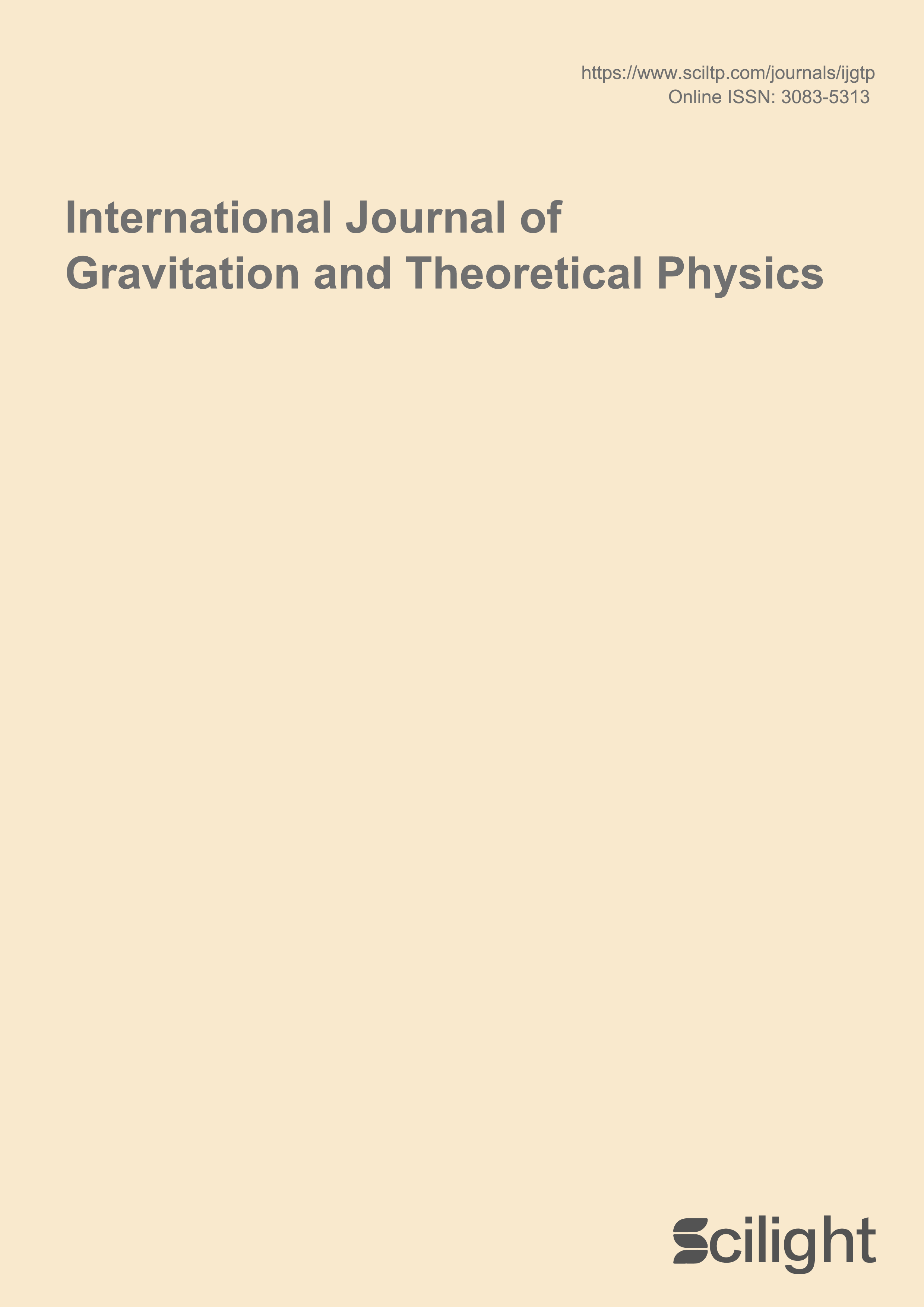} \vspace{-3pt}\\
        \end{tabular}
        \vspace{-22pt}
    \end{table}}
   
    \fancyfoot[C]{
        \vspace{-1.55cm}
        \begin{table}[H]
            \begin{minipage}[c]{0.15\columnwidth}
                \includegraphics[scale=0.5]{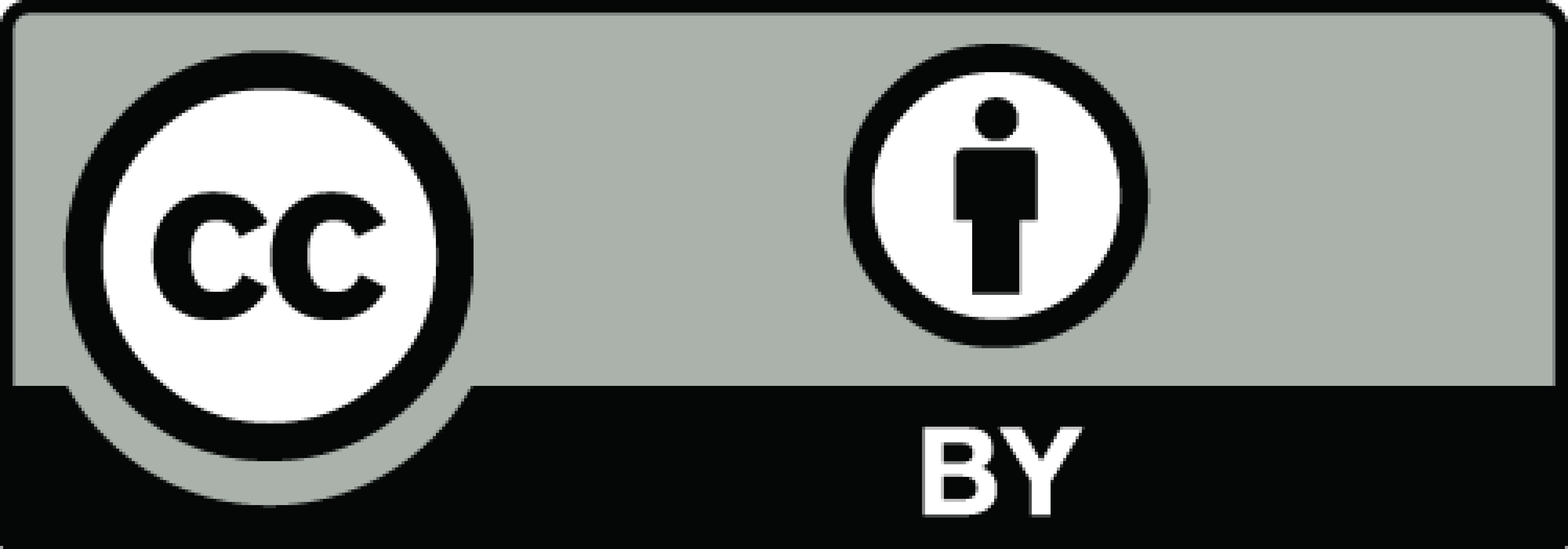} \vspace{1.1pt}
            \end{minipage}
            \hfill
            \begin{minipage}[c]{0.85\columnwidth}
                \scriptsize \textbf{Copyright:} © 2026 by the authors. This is an open access article under the terms and conditions of the Creative Commons Attribution (\mbox{CC BY}) license  (\href{https://creativecommons.org/licenses/by/4.0/}{https://creativecommons.org/licenses/by/4.0/}). \\ \textbf{Publisher’s Note:} Scilight stays neutral with regard to jurisdictional claims in published maps and institutional affiliations.
            \end{minipage}
    \end{table}}
    \vspace{-0.55cm}
}

\begin{document}
\newgeometry{left=2.5cm, right=2.5cm, top=1.8cm, bottom=4cm}
	\thispagestyle{firstpage}
	\nolinenumbers  
   {\noindent \textit{{Article
}} 
\hfill
\href{https://crossmark.crossref.org/dialog/?doi=10.53941/ams.2026.100010\&domain=pdf}{
    \includegraphics[height=1.3em]{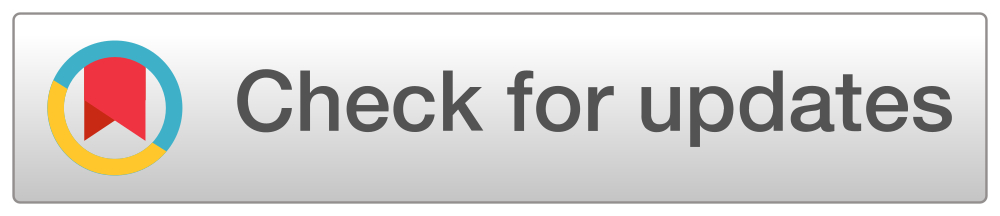}
    \hspace*{-0.28cm}
  }}
    \vspace{4pt} \\
	{\fontsize{18pt}{10pt}\textbf{Hawking Temperatures and Radiation Estimates for Dilaton--de Sitter Black Holes}  }
	\vspace{16pt} \\
	{\large Can Ertugay}
	\vspace{6pt}
	 \begin{spacing}{0.9}
		{\noindent \small
				\parbox[t]{0.98\linewidth}{Physics Department, Faculty of Science, Akdeniz University, Dumlupinar Boulevard, Campus, Antalya 07058 
, Turkiye; canertugay@gmail.com} \vspace{6pt}\\
		\footnotesize	\textbf{How To Cite}: Ertugay, C. Hawking Temperatures and Radiation Estimates for Dilaton--de Sitter Black Holes. \emph{International Journal of Gravitation and Theoretical Physics} \textbf{2026}, \emph{2}(2), 2. \href{https://doi.org/10.53941/ijgtp.2026.200002}{https://doi.org/10.53941/ijgtp.2026.200002}}\\
	\end{spacing}

\begin{table}[H]
\noindent\rule[0.15\baselineskip]{\textwidth}{0.5pt} 
\begin{tabular}{lp{12cm}}  
 \small 
  \begin{tabular}[t]{@{}l@{}} 
  \footnotesize  Received: 24 May 2026 \\
  \footnotesize  Revised: 17 June 2026 \\
   \footnotesize Accepted: 18 June 2026 \\
  \footnotesize  Published: 23 June 2026
  \end{tabular} &
  \textbf{Abstract:} Charged dilaton black holes with a positive cosmological constant provide a useful arena in which to test how scalar hair modifies semiclassical physics in a spacetime with two Killing horizons. The Gao--Zhang solution realizes such a geometry in Einstein--Maxwell--dilaton theory by replacing a single Liouville potential, which is insufficient for asymptotically de Sitter boundary conditions, by a three-Liouville dilaton potential. Although the solution and several of its perturbative and optical properties have been studied, its temperature and heat capacity have not been examined through the same range of temperature prescriptions commonly considered for Schwarzschild--de Sitter black holes, where the absence of global thermal equilibrium motivates several inequivalent temperature definitions. We present this temperature-prescription comparison for the four-dimensional, string-coupling member of the Gao--Zhang family. We compare the standard surface-gravity temperature, the Bousso--Hawking-normalized temperature, and two effective temperatures built from the black-hole and cosmological horizons. The dilaton changes the areal radius, entropy, photon-sphere condition, and greybody problem, while the de Sitter normalization ambiguity changes radiation estimates by powers of the redshift factor. For representative parameters, the Bousso--Hawking prescription can enhance a Stefan--Boltzmann estimate of the black-hole power by one to two orders of magnitude relative to the unnormalized surface-gravity prescription, whereas the entropy-sum effective temperature can suppress the same estimate. These results identify a concrete gap in the thermodynamics of dilaton--de Sitter black holes and provide a roadmap for a full greybody-factor calculation. \\
\\
  & 

\end{tabular}
\noindent\rule[0.15\baselineskip]{\textwidth}{0.5pt} 
\end{table}

	\section{Introduction}

\textls[-15]{Dilaton fields arise naturally in low-energy string theory and in many effective descriptions of compactified higher-dimensional gravity. Once the dilaton couples nonminimally to the Maxwell sector, the causal structure and thermodynamics of charged black holes are changed in ways that are already visible in asymptotically flat Einstein--Maxwell--dilaton solutions such as the Gibbons--Maeda and Garfinkle--Horowitz--Strominger black holes \cite{GibbonsMaeda1988,GarfinkleHorowitzStrominger1991}.} In particular, the surface area of symmetry spheres is no longer simply $4\pi r^2$, the curvature singularity may be shifted to a finite positive value of the radial coordinate, and the electric charge is tied to scalar hair.

The inclusion of a cosmological constant is subtler. In a simple Liouville-potential model, no static, spherically symmetric, charged dilaton black hole with ordinary de Sitter or anti-de Sitter asymptotics exists, apart from the trivial pure-cosmological-constant case. Gao and Zhang overcame this obstruction in four dimensions by introducing a potential consisting of three Liouville terms and constructed exact charged dilaton black holes in de Sitter and anti-de Sitter backgrounds \cite{GaoZhang2004}. Higher-dimensional analogues were subsequently obtained \cite{Dehghani2005}, and related solutions have been used as backgrounds for particle motion and strong lensing \cite{MukherjeeMajumdar2007}, \textls[-15]{quasinormal modes \cite{Fernando2016,DubinskyZinhailo2024}, thermodynamics of AdS dilaton branes \cite{Sheykhi2007,SheykhiDehghaniHendi2010},} and generalized Einstein--Maxwell--scalar constructions with first-law and Smarr analyses \cite{YuQiuGao2021}. A more recent de Sitter thermodynamic treatment of topological dilaton black holes with nonlinear electrodynamics also emphasizes that the black-hole and cosmological horizons are thermodynamically linked \cite{ZhaoZhangGaoLiu2021}.

There remains, however, a specific missing piece. In de Sitter black-hole thermodynamics the timelike Killing vector is not normalized at spatial infinity, because there is no spatial infinity inside the static patch. The ``temperature of the black hole'' is therefore not a unique object. For Schwarzschild--de Sitter black holes, this issue has led to several prescriptions: the bare surface-gravity temperature, the Bousso--Hawking temperature obtained by normalizing the Killing vector at the geodesic static radius, and effective temperatures constructed from both the black-hole and cosmological horizons \cite{BoussoHawking1996,BoussoHawking1998,KantiPappas2017,Pappas:2017kam,Pappas:2016ovo}. Kanti and Pappas showed that these choices can lead to materially different Hawking spectra and even different conclusions about relative emissivities \cite{KantiPappas2017}. The analogous comparison has not, to our knowledge, been carried out for the Gao--Zhang dilaton--de Sitter black holes.

This paper formulates that comparison. We focus on the four-dimensional string-coupling case, for which the metric takes a compact form and is the case most often used in perturbative studies. The generalization to arbitrary dilaton coupling is straightforward at the level of principle but less transparent algebraically. Section~\ref{sec:theory} reviews the action, the solution, the horizon structure, and the entropy. Section~\ref{sec:temperatures} introduces the different temperature prescriptions in a form adapted to the dilaton geometry. Section~\ref{sec:thermo} discusses thermodynamic consequences, including heat-capacity diagnostics and the role of the cosmological horizon. Section~\ref{sec:radiation} estimates Hawking emission rates and shows how strongly the predicted flux depends on the chosen de Sitter temperature. We work in units $G=c=\hbar=k_{\rm B}=1$.

\section{Theory and Dilaton--de Sitter Geometry}\label{sec:theory}

\subsection{Einstein--Maxwell--Dilaton Action}

The four-dimensional Einstein--Maxwell--dilaton model may be written as
\begin{equation}
  I=\frac{1}{16\pi}\int \dd^4x\sqrt{-g}\left[
  R-2(\nabla\Phi)^2-V(\Phi)-\e^{-2\alpha\Phi}F_{\mu\nu}F^{\mu\nu}
  \right].
  \label{eq:action_general}
\end{equation}

Here $\alpha$ controls the coupling between the scalar and electromagnetic fields. A single Liouville term in $V(\Phi)$ does not generate the desired de Sitter or anti-de Sitter asymptotics for nonzero charge and nontrivial dilaton. Gao and Zhang instead used a three-Liouville potential. In a common normalization, and setting an additive constant in $\Phi$ to zero for notational simplicity, it can be expressed as
\begin{equation}
  V(\Phi)=\frac{2\Lambda}{3(1+\alpha^2)^2}
  \left[\alpha^2(3\alpha^2-1)\e^{-2\Phi/\alpha}
  +(3-\alpha^2)\e^{2\alpha\Phi}
  +8\alpha^2\e^{(\alpha-\alpha^{-1})\Phi}\right].
  \label{eq:three_liouville}
\end{equation}

For $\Phi=0$, this reduces to $V(0)=2\Lambda$, so the action reproduces the Einstein--Maxwell theory with cosmological constant in the nondilatonic limit.

The detailed analysis below specializes to the string value $\alpha=1$. The potential is then
\begin{equation}
  V(\Phi)=\frac{\Lambda}{3}\left(\e^{2\Phi}+\e^{-2\Phi}+4\right),
  \label{eq:potential_alpha_one}
\end{equation}
up to the harmless shift $\Phi\rightarrow \Phi-\Phi_0$. The field equations following from Equation~\eqref{eq:action_general} are
\begin{align}
  R_{\mu\nu}&=2\partial_\mu\Phi\partial_\nu\Phi
  +\frac{1}{2}g_{\mu\nu}V(\Phi)
  +2\e^{-2\Phi}\left(F_{\mu\lambda}F_{\nu}{}^{\lambda}
  -\frac{1}{4}g_{\mu\nu}F^2\right),\label{eq:einstein_eq}\\
  \nabla_\mu\left(\e^{-2\Phi}F^{\mu\nu}\right)&=0,\label{eq:maxwell_eq}\\
  \nabla^2\Phi&=\frac{1}{4}\frac{\dd V}{\dd\Phi}-\frac{1}{2}\e^{-2\Phi}F^2.\label{eq:dilaton_eq}
\end{align}

\subsection{Metric, Fields, and Limits}

For $\Lambda>0$ the four-dimensional charged dilaton--de Sitter metric can be written as
\begin{equation}
  \dd s^2=-f(r)\dd t^2+\frac{\dd r^2}{f(r)}+R^2(r)\dd\Omega_2^2,
  \label{eq:metric}
\end{equation}
with

\begin{align}
  f(r)&=1-\frac{2M}{r}-\frac{\Lambda}{3}r(r-2Q),\label{eq:f}\\
  R^2(r)&=r(r-2Q),\label{eq:areal_radius}\\
  \e^{2\Phi(r)}&=\e^{2\Phi_0}\left(1-\frac{2Q}{r}\right),\label{eq:dilaton}\\
  F_{tr}&=\frac{q\e^{2\Phi_0}}{r^2},\qquad
  Q=\frac{q^2\e^{2\Phi_0}}{2M}.
  \label{eq:maxwell_solution}
\end{align}

The coordinate $r$ is not the areal radius; the area of a two-sphere at coordinate radius $r$ is $4\pi R^2(r)$. The physical static region must lie outside the dilatonic singular surface
\begin{equation}
  r_s=2Q,
  \label{eq:singularity}
\end{equation}
where the two-sphere shrinks and curvature invariants diverge. Thus a regular black-hole exterior requires the black-hole horizon $r_\bhor$ to satisfy $r_\bhor>2Q$.

The two limiting cases are instructive. Setting $Q=0$ removes the scalar hair and Equation~\eqref{eq:f} reduces to Schwarzschild--de Sitter. Setting $\Lambda=0$ gives the electrically charged stringy dilaton black hole, with event horizon at $r=2M$ and singular surface at $r=2Q$. The de Sitter case combines both features: a cosmological horizon appears, but the black-hole area and entropy retain explicit dilaton dependence.

\subsection{Horizons and Nariai-Type Endpoint}

Killing horizons are roots of $f(r)=0$. Multiplying Equation~\eqref{eq:f} by $r$ gives the cubic
\begin{equation}
  -\frac{\Lambda}{3}r^3+\frac{2\Lambda Q}{3}r^2+r-2M=0.
  \label{eq:horizon_cubic}
\end{equation}

For appropriate $(M,Q,\Lambda)$ there are two positive roots outside $r_s=2Q$; these are the black-hole horizon $r_\bhor$ and the cosmological horizon $r_\chor$, with
\begin{equation}
  2Q<r_\bhor<r_\chor.
  \label{eq:horizon_ordering}
\end{equation}

Throughout the paper $r_h$ denotes a generic Killing-horizon radius. It can be replaced by $r_\bhor$ when the formula is evaluated at the black-hole horizon, or by $r_\chor$ when it is evaluated at the cosmological horizon. Thus $r_h$ is not a third radius or an additional horizon; it is a placeholder used for expressions, such as the entropy or surface gravity, that have the same form at either root of $f(r)=0$. By contrast, $r_\bhor$ always refers specifically to the smaller positive root outside the singular surface and determines the black-hole area, temperature, and emitting surface.
The mass parameter may be expressed in terms of any horizon radius $r_h$ as
\begin{equation}
  M(r_h)=\frac{r_h}{2}\left[1-\frac{\Lambda}{3}r_h(r_h-2Q)\right].
  \label{eq:mass_horizon}
\end{equation}

The de Sitter black-hole and cosmological horizons coalesce when $f(r_N)=f'(r_N)=0$. For the present metric this gives
\begin{align}
  r_N&=\frac{2Q}{3}+\sqrt{\frac{4Q^2}{9}+\frac{1}{\Lambda}},\label{eq:nariai_radius}\\
  M_N&=\frac{\Lambda}{3}r_N^2(r_N-Q).
  \label{eq:nariai_mass}
\end{align}

For $Q=0$ these formulas reduce to the usual Schwarzschild--de Sitter values $r_N=\Lambda^{-1/2}$ and $M_N=(3\sqrt{\Lambda})^{-1}$.

The horizon entropy is one quarter of the area,
\begin{equation}
  S_h=\frac{A_h}{4}=\pi R^2(r_h)=\pi r_h(r_h-2Q),
  \label{eq:entropy}
\end{equation}
which vanishes as a horizon approaches the dilatonic singular surface. The electric potential measured relative to a horizon may be chosen, up to a gauge constant, as
\begin{equation}
  \Phi_e(r_h)\simeq \frac{q\e^{2\Phi_0}}{r_h},
  \label{eq:electric_potential}
\end{equation}
with the caveat that in de Sitter space thermodynamic potentials are usually defined quasilocally or by comparing the two horizons.

\section{Hawking Temperature Prescriptions}\label{sec:temperatures}

\subsection{Bare Surface-Gravity Temperature}

For the Killing vector $\chi=\partial_t$, the standard horizon surface gravity is
\begin{equation}
  \kappa_h^{(0)}=\frac{1}{2}|f'(r_h)|,
  \label{eq:surface_gravity}
\end{equation}
so that the corresponding ``bare'' temperature is
\begin{equation}
  T_h^{(0)}=\frac{|f'(r_h)|}{4\pi},\qquad
  f'(r)=\frac{2M}{r^2}-\frac{2\Lambda}{3}(r-Q).
  \label{eq:bare_temp}
\end{equation}

Using the horizon equation to eliminate $M$, one obtains
\begin{align}
  T_\bhor^{(0)}&=\frac{1}{4\pi}\left(\frac{1}{r_\bhor}-\Lambda r_\bhor+\frac{4\Lambda Q}{3}\right),\label{eq:Tbare_b}\\
  T_\chor^{(0)}&=\frac{1}{4\pi}\left(\Lambda r_\chor-\frac{1}{r_\chor}-\frac{4\Lambda Q}{3}\right),\label{eq:Tbare_c}
\end{align}
where the signs are chosen so that both temperatures are positive in the static region. These temperatures vanish at the Nariai point. Away from special parameter values, $T_\bhor^{(0)}\neq T_\chor^{(0)}$, so the static patch is not in ordinary thermal equilibrium.

\subsection{Tolman Temperature and Bousso--Hawking Normalization}

Because the spacetime is not asymptotically flat, the normalization of $\chi$ is ambiguous. A static observer at radius $r$ measures the Tolman-redshifted temperature
\begin{equation}
  T_{h,\loc}^{(0)}(r)=\frac{T_h^{(0)}}{\sqrt{f(r)}}.
  \label{eq:tolman_temp}
\end{equation}

\textls[-15]{The Bousso--Hawking prescription chooses a preferred observer at the point where the gravitational attraction of the black hole and the cosmological repulsion balance. In the present geometry this geodesic radius $r_0$ is defined by}
\begin{equation}
  f'(r_0)=0,
  \qquad
  r_0^3-Qr_0^2-\frac{3M}{\Lambda}=0,
  \qquad
  r_\bhor<r_0<r_\chor.
  \label{eq:r0}
\end{equation}

The normalized black-hole temperature is then
\begin{equation}
  T_\bhor^{\BH}=\frac{T_\bhor^{(0)}}{\sqrt{f(r_0)}}.
  \label{eq:TBH}
\end{equation}

One may analogously normalize $T_\chor^{(0)}$, although much of the Schwarzschild--de Sitter radiation literature uses $T_\bhor^{\BH}$ together with the bare cosmological-horizon temperature when forming effective quantities. The normalization factor is physically important: as the horizons approach the Nariai limit, both $T_\bhor^{(0)}$ and $\sqrt{f(r_0)}$ vanish, and their ratio can remain finite.

\subsection{Effective Two-Horizon Temperatures}

The static patch contains two horizons with distinct entropies. This motivates effective temperatures that encode the response of a combined horizon system. If $T_\bhor$ denotes the black-hole temperature used in the prescription and $T_\chor\equiv T_\chor^{(0)}$, two commonly used combinations are
\begin{align}
  T_{\eff}^{(+)}&=\frac{T_\bhor T_\chor}{T_\bhor+T_\chor},\label{eq:Teff_plus}\\
  T_{\eff}^{(-)}&=\frac{T_\bhor T_\chor}{T_\bhor-T_\chor}.
  \label{eq:Teff_minus}
\end{align}

The plus prescription is associated with adding horizon entropies, while the minus prescription appears when the energy flow through one horizon is assigned the opposite sign to the flow through the other. A Bousso--Hawking effective temperature can be defined by substituting $T_\bhor=T_\bhor^{\BH}$ in Equation~\eqref{eq:Teff_minus}:
\begin{equation}
  T_{\eff}^{\BH}=\frac{T_\bhor^{\BH}T_\chor^{(0)}}{T_\bhor^{\BH}-T_\chor^{(0)}}.
  \label{eq:Teff_BH}
\end{equation}

Equations~\eqref{eq:TBH}--\eqref{eq:Teff_BH} are the direct dilaton--de Sitter analogues of the temperature prescriptions compared for Schwarzschild--de Sitter radiation spectra. The new ingredient is that $Q$ affects the horizons, the normalization point, and the entropy through the areal radius $R(r)$.

\section{Thermodynamic Consequences}\label{sec:thermo}

\subsection{Entropy and First-Law Bookkeeping}

The individual horizon entropies are
\begin{equation}
  S_\bhor=\pi r_\bhor(r_\bhor-2Q),
  \qquad
  S_\chor=\pi r_\chor(r_\chor-2Q).
  \label{eq:two_entropies}
\end{equation}

Thus the dilaton shifts both the entropy and its variation,
\begin{equation}
  \dd S_h=2\pi(r_h-Q)\dd r_h-2\pi r_h\dd Q.
  \label{eq:entropy_variation}
\end{equation}

This is already enough to show why a careful thermodynamic treatment cannot be obtained by copying the Schwarzschild--de Sitter formulae with $r_h^2$ left unchanged. If one varies the physical electric charge $q$, then $Q=q^2\e^{2\Phi_0}/(2M)$ also varies; if instead one holds $Q$ fixed, one is working in a useful but not fully canonical slice of parameter space. A complete first law should therefore keep track of the electric work term and the scalar contribution, or else use a quasilocal formalism adapted to the static patch.

For the purpose of isolating the temperature ambiguity, it is useful to consider fixed $Q$ and $\Lambda$ and use Equation~\eqref{eq:mass_horizon}. Then
\begin{equation}
  \frac{\dd M}{\dd r_\bhor}
  =\frac{1}{2}\left(1-\Lambda r_\bhor^2+\frac{4\Lambda Q r_\bhor}{3}\right)
  =\frac{r_\bhor}{2}f'(r_\bhor).
  \label{eq:dMdr}
\end{equation}

The unnormalized heat-capacity diagnostic is
\begin{equation}
  C_0\equiv\left(\frac{\partial M}{\partial T_\bhor^{(0)}}\right)_{Q,\Lambda}
  =-\frac{2\pi r_\bhor^2\left(1-\Lambda r_\bhor^2+4\Lambda Q r_\bhor/3\right)}{1+\Lambda r_\bhor^2}.
  \label{eq:C0}
\end{equation}

In the ordinary black-hole branch, where $T_\bhor^{(0)}>0$, this diagnostic is negative. Other prescriptions replace $T_\bhor^{(0)}$ by $T_\bhor^{\BH}$ or by an effective two-horizon temperature; because $r_0$ and $r_\chor$ vary with $M$, their heat capacities can develop different zeros and poles. Thus even before computing greybody factors, the choice of de Sitter temperature changes the inferred stability properties.

\subsection{Lukewarm and Near-Nariai Regimes}

Equilibrium between the two bare horizons would require
\begin{equation}
  T_\bhor^{(0)}=T_\chor^{(0)},
  \qquad
  \frac{1}{r_\bhor}+\frac{1}{r_\chor}
  =\Lambda\left(r_\bhor+r_\chor-\frac{8Q}{3}\right).
  \label{eq:lukewarm_condition}
\end{equation}

This condition is not generic. The dilaton charge shifts the balance relative to Reissner--Nordström--de Sitter or Schwarzschild--de Sitter black holes, and it should be tested against the requirement $r_\bhor>2Q$.

Near the Nariai point, Equation~\eqref{eq:nariai_radius}, the bare temperatures tend to zero. In contrast, $T_\bhor^{\BH}$ need not vanish because the normalization point also approaches a zero of $f(r)$. This is precisely the type of regime in which the Bousso--Hawking normalization can qualitatively alter evaporation estimates. Since the dilaton reduces the areal radius according to $R^2=r(r-2Q)$, the near-Nariai entropy and luminosity are additionally suppressed by the scalar hair even when the normalized temperature remains finite.

\section{Hawking-Radiation Estimates}\label{sec:radiation}

\subsection{Scalar Wave Equation and Greybody Factors}

A massless scalar perturbation $\varphi$ obeys
\begin{equation}
  \Box\varphi=\xi {\cal R},
  \label{eq:wave_eq}
\end{equation}
where $\xi$ is the coupling parameter. In particular, $\xi=1/6$ corresponds to the conformal coupling.

With the separation (see details in \cite{Carter:1968ks,Konoplya:2018arm})
\begin{equation}
  \varphi=\frac{\psi_{\omega\ell}(r)}{R(r)}Y_{\ell m}(\theta,\phi)\e^{-i\omega t},
  \qquad
  \frac{\dd r_*}{\dd r}=\frac{1}{f(r)},
  \label{eq:separation}
\end{equation}
the radial equation takes the Schrodinger form
\begin{equation}
  \frac{\dd^2\psi_{\omega\ell}}{\dd r_*^2}
  +\left[\omega^2-V_\ell(r)\right]\psi_{\omega\ell}=0,
  \label{eq:radial_wave}
\end{equation}
where the effective potential may be written as
\begin{equation}
  V_\ell(r)=f(r)\left[\frac{\ell(\ell+1)}{R^2(r)}+\frac{1}{R(r)}\frac{\dd}{\dd r}\left(f(r)\frac{\dd R}{\dd r}\right)-
        \xi{\cal R}(r)\right],
  \label{eq:scalar_potential}
\end{equation}
where
\[
{\cal R}(r)=\frac{2Q^2 f(r)}{r^2(r-2Q)^2}+\frac{2\Lambda}{3}\left[
\frac{r-2Q}{r}+\frac{r}{r-2Q}+4\right].
\]

The greybody factor $\Gamma_\ell(\omega)$ is obtained by solving Equation~\eqref{eq:radial_wave} with the appropriate ingoing and outgoing boundary conditions at the two horizons. The corresponding energy-emission spectrum for a temperature choice $T_X$ is
\begin{equation}
  \frac{\dd^2E_X}{\dd t\,\dd\omega}
  =\frac{1}{2\pi}\sum_{\ell=0}^{\infty}(2\ell+1)
  \frac{\Gamma_\ell(\omega)\,\omega}{\exp(\omega/T_X)-1}.
  \label{eq:spectrum}
\end{equation}

Equation~\eqref{eq:spectrum} shows why the de Sitter temperature ambiguity matters: even if the same greybody factors are used, the Planck denominator changes nonlinearly with $T_X$.

At high frequencies, a useful geometric-optics estimate is obtained from the photon sphere. For the metric \eqref{eq:metric}, null circular orbits satisfy
\begin{equation}
  \frac{\dd}{\dd r}\left(\frac{R^2(r)}{f(r)}\right)_{r=r_{\rm ph}}=0,
  \qquad
  \frac{f'(r_{\rm ph})}{f(r_{\rm ph})}
  =\frac{2(r_{\rm ph}-Q)}{r_{\rm ph}(r_{\rm ph}-2Q)}.
  \label{eq:photon_sphere}
\end{equation}

The associated critical impact parameter is
\begin{equation}
  b_{\rm ph}^2=\frac{R^2(r_{\rm ph})}{f(r_{\rm ph})},
  \label{eq:impact_parameter}
\end{equation}
so a geometric absorption cross section is $\sigma_{\rm geo}\simeq\pi b_{\rm ph}^2$.

At low frequencies the behavior is qualitatively different from the asymptotically flat case. Because the wave is scattered between two horizons rather than between a horizon and spatial infinity, the minimally coupled $s$ wave can approach a nonzero transmission probability as $\omega\rightarrow0$. This constant can be obtained by a standard matched zero-frequency argument. For $\ell=0$ it is useful to work with the unrescaled radial mode $\phi_0(r)$, whose equation is
\begin{equation}
  \frac{\dd}{\dd r}\left[R^2(r)f(r)\frac{\dd\phi_0}{\dd r}\right]
  +\frac{\omega^2R^2(r)}{f(r)}\phi_0-\xi{\cal R}(r)R^2(r)\phi_0=0.
  \label{eq:s_wave_radial}
\end{equation}

At zero frequency ($\omega=0$) and coupling ($\xi=0$) this equation integrates to
\begin{equation}
  R^2(r)f(r)\frac{\dd\phi_0}{\dd r}=C_2,\qquad
  \phi_0=C_1+C_2\int^r\frac{\dd r'}{R^2(r')f(r')}.
  \label{eq:zero_frequency_solution}
\end{equation}

Near either horizon, $\dd r_*/\dd r=1/f$ implies
\begin{equation}
  \int^r\frac{\dd r'}{R^2(r')f(r')}
  \simeq \frac{r_*}{R^2(r_h)},
  \label{eq:horizon_zero_frequency}
\end{equation}
so the same integration constant $C_2$ corresponds to different slopes in $r_*$, weighted by the two horizon areas. The low-frequency scattering solution has the asymptotic form
\begin{equation}
  \phi_0\simeq A_{\rm tr}e^{-i\omega r_*}\quad (r\to r_\bhor),\qquad
  \phi_0\simeq A_{\rm in}e^{-i\omega r_*}+A_{\rm out}e^{i\omega r_*}\quad (r\to r_\chor),
  \label{eq:low_frequency_scattering}
\end{equation}
where the wave is purely ingoing at the black-hole horizon and is a superposition of incident and reflected waves at the cosmological horizon. Expanding Equation~\eqref{eq:low_frequency_scattering} for small $\omega$ and matching the constant and linear terms in $r_*$ to Equation~\eqref{eq:horizon_zero_frequency} for the minimal coupling ($\xi=0$) gives
\begin{equation}
  A_{\rm tr}=A_{\rm in}+A_{\rm out},\qquad
  \frac{A_{\rm tr}R^2(r_\bhor)}{R^2(r_\chor)}=A_{\rm in}-A_{\rm out}.
  \label{eq:matching_amplitudes}
\end{equation}

Therefore
\begin{equation}
  \frac{A_{\rm tr}}{A_{\rm in}}=
  \frac{2R^2(r_\chor)}{R^2(r_\bhor)+R^2(r_\chor)}.
  \label{eq:transmission_amplitude}
\end{equation}

The radial flux carried by a horizon plane wave is proportional to the area factor $R^2(r_h)$ times the squared amplitude. The resulting zero-frequency greybody factor is therefore (cf.~\cite{Kanti:2005ja})
\begin{equation}
  \Gamma_0(0)=\frac{\mathcal F_{\rm tr}}{\mathcal F_{\rm in}}\simeq
  \frac{R^2(r_\bhor)}{R^2(r_\chor)}
  \left|\frac{A_{\rm tr}}{A_{\rm in}}\right|^2
  =\frac{4R^2(r_\bhor)R^2(r_\chor)}{\left[R^2(r_\bhor)+R^2(r_\chor)\right]^2},
  \label{eq:low_frequency_greybody}
\end{equation}
which reduces to the usual Schwarzschild--de Sitter area formula when $Q=0$. The dilaton enters this estimate through the replacement of the areal factors $r_h^2$ by $R^2(r_h)=r_h(r_h-2Q)$. Hence scalar hair suppresses the low-frequency transmission whenever it reduces the black-hole area relative to the cosmological area, and the suppression becomes strong as $r_\bhor$ approaches the singular surface $2Q$.

This result also clarifies the meaning of a low-frequency absorption cross section in a de Sitter static patch. If one formally uses the partial-wave relation
\begin{equation}
  \sigma_{\rm abs}(\omega)=\frac{\pi}{\omega^2}
  \sum_{\ell=0}^{\infty}(2\ell+1)\Gamma_\ell(\omega),
  \label{eq:partial_wave_cross_section}
\end{equation}
then the $s$-wave term behaves as $\sigma_{\rm abs}\sim\pi\Gamma_0(0)/\omega^2$ rather than approaching the black-hole horizon area. This infrared growth reflects the finite cosmological horizon boundary condition and should not be confused with the asymptotically flat universality result $\sigma_{\rm abs}(0)=A_\bhor$. In practical luminosity estimates the relevant quantity is therefore the finite transmission probability in Equation~\eqref{eq:low_frequency_greybody}, together with the emitting area or the geometric-optics cross section used to normalize the flux. 

However, for the nonminimally coupled scalar field in the asymptotically de Sitter spacetime it is possible to show that the greybody factor for the $\ell=0$ mode generally vanishes in the zero-frequency limit 
 \cite{Crispino:2013pya}. The coupling term in Equation~\eqref{eq:s_wave_radial} leads to the nontrivial logarithmic correction, and the matching conditions read:
\begin{equation}
    A_{\rm in}+A_{\rm out}=A_{\rm tr}, \qquad
    i\omega(A_{\rm out}-A_{\rm in})\simeq A_{\rm tr}B,
\end{equation}
where
\begin{equation}
    B=
    \frac{\xi}{R^2(r_c)}
    \int_{r_\bhor}^{r_c}
    {\cal R}(r)R^2(r)\phi_0(r)\,dr .
    \label{eq:B_def}
\end{equation}

Although we do not give an explicit form of the integral in~\eqref{eq:B_def}, it is finite and can be obtained perturbatively in $\xi$ by solving the corresponding Volterra integral equation.

The greybody factor in the zero-frequency limit approaches zero as
\begin{equation}
  \Gamma_0(\omega)=
  \frac{R^2(r_\bhor)}{R^2(r_\chor)}
  \left|\frac{A_{\rm tr}}{A_{\rm in}}\right|^2
  =\frac{4R^2(r_\bhor)}{R^2(r_c)|B|^2}\,\omega^2+\mathcal{O}(\omega)^2.
  \label{eq:low_frequency_greybody_coupling}
\end{equation}

Consequently the $s$-wave absorption cross section is finite in the infrared limit,
\begin{equation}
\sigma_{\rm abs}^{(0)}(0)
=\frac{4\pi R^2(r_\bhor)}{R^2(r_c)|B|^2}
=\frac{4\pi R^2(r_\bhor) R^2(r_c)}{\xi^2
\left|\displaystyle
\int_{r_\bhor}^{r_c}{\cal R}(r)R^2(r)\phi_0(r)\,dr
\right|^2
}.
\label{eq:sigma_nonminimal_finite}
\end{equation}

A more precise frequency-dependent cross section for the dilaton geometry requires solving Equation~\eqref{eq:radial_wave} in the whole range of frequencies either numerically \cite{Page:1976ki,Page:1976df,Calza:2025whq} or semi-analytically \cite{Konoplya:2023moy,Iyer:1986np}.

Calculations of greybody factors with the 6th order WKB accuracy for a scalar and Dirac fields have been done in \cite{Lutfuoglu:2025eik}. However, the WKB accuracy is usually insufficient for sufficiently accurate estimates of the Hawking radiation \cite{Tan:2026vif,Lutfuoglu:2026gey,Lutfuoglu:2025hjy,Bolokhov:2026uol,Bolokhov:2026kqu,Skvortsova:2024msa}.

\subsection{Stefan--Boltzmann Estimates}

\textls[-15]{A compact way to compare temperature prescriptions is to use a greybody-averaged Stefan--Boltzmann estimate}
\begin{equation}
  P_X\simeq g_*\,\sigma_{\rm SB}\,A_{\rm eff}\,T_X^4,
  \qquad
  \sigma_{\rm SB}=\frac{\pi^2}{60},
  \label{eq:SB}
\end{equation}
where $g_*$ counts effectively massless degrees of freedom and $A_{\rm eff}$ is either the horizon area $4\pi R^2(r_\bhor)$ or the geometric-optics area $\pi b_{\rm ph}^2$. The absolute value of $P_X$ is model-dependent, but ratios of estimates based on the same $A_{\rm eff}$ obey
\begin{equation}
  \frac{P_X}{P_0}\simeq \left(\frac{T_X}{T_\bhor^{(0)}}\right)^4.
  \label{eq:power_ratio}
\end{equation}

Thus the Bousso--Hawking prescription gives
\begin{equation}
  \frac{P_{\BH}}{P_0}\simeq \frac{1}{f^2(r_0)},
  \label{eq:PBH_ratio}
\end{equation}
which can be large when $f(r_0)$ is small. The effective temperatures instead may either suppress or enhance the power depending on whether $T_{\eff}$ lies below or above $T_\bhor^{(0)}$.

Table~\ref{tab:benchmark} gives representative dimensionless estimates with $M=1$, $Q=1/2$, and three values of $\Lambda$. The singular surface is at $r_s=1$, so the listed black-hole horizons cloak the singularity. The table is not a substitute for a greybody computation; it isolates the temperature effect that any such computation would inherit.
\vspace{-12pt}

\begin{table}[H]
\centering
\small
\caption{Temperature estimates for the dilaton--de Sitter black hole with $M=1$ and $Q=1/2$. The last column uses Equation~\eqref{eq:PBH_ratio}. The entropy-sum effective temperature $T_{\eff}^{(+)}$ is smaller than the bare black-hole temperature in all three examples, while $T_\bhor^{\BH}$ is larger because the preferred static observer sits at a finite redshift.}
\label{tab:benchmark}
\newcolumntype{c}{>{\centering\arraybackslash}X}
\begin{tabularx}{\textwidth}{ccccccccc} 
\toprule
$\bm{\Lambda}$ & $\bm{r_\bhor}$ & $\bm{r_\chor}$ & $\bm{f(r_0)}$ & $\bm{T_\bhor^{(0)}}$ & $\bm{T_\bhor^{\BH}}$ & $\bm{T_\chor^{(0)}}$ & $\bm{T_{\eff}^{(+)}}$ & $\bm{P_{\BH}/P_0}$ \\
\midrule
$0.02$ & $2.028$ & $11.659$ & $0.471$ & $0.0371$ & $0.0540$ & $0.0107$ & $0.0083$ & $4.50$ \\
$0.08$ & $2.139$ & $5.379$  & $0.195$ & $0.0278$ & $0.0630$ & $0.0152$ & $0.0098$ & $26.2$ \\
$0.12$ & $2.255$ & $4.122$  & $0.094$ & $0.0201$ & $0.0654$ & $0.0137$ & $0.0082$ & $112$ \\
\bottomrule
\end{tabularx}

\end{table}

The numerical trend is clear. Increasing $\Lambda$ pushes the two horizons together, lowers the bare black-hole temperature, and decreases $f(r_0)$. The Bousso--Hawking temperature grows relative to the bare temperature and can therefore raise a $T^4$ luminosity estimate by orders of magnitude. Conversely, $T_{\eff}^{(+)}$ remains tied to the colder cosmological horizon and suppresses the black-hole emission estimate. The minus prescription $T_{\eff}^{(-)}$ behaves differently: it can become large when $T_\bhor^{(0)}\approx T_\chor^{(0)}$, signalling that this effective temperature should be interpreted as a thermodynamic response function rather than as an ordinary local bath temperature.

\subsection{Evaporation Direction and Charge Dependence}

At fixed $Q$, the bare black-hole temperature decreases as the black-hole horizon grows,
\begin{equation}
  \frac{\partial T_\bhor^{(0)}}{\partial r_\bhor}
  =-\frac{1}{4\pi}\left(\frac{1}{r_\bhor^2}+\Lambda\right)<0.
  \label{eq:dTdr}
\end{equation}

The dilaton charge also reduces the emitting area,
\begin{equation}
  A_\bhor=4\pi r_\bhor(r_\bhor-2Q),
  \label{eq:emitting_area}
\end{equation}
which suppresses the luminosity as $Q$ approaches $r_\bhor/2$. These two effects compete with the normalization effect in Equation~\eqref{eq:TBH}. A full evaporation model should therefore evolve at least $(M,q)$ rather than $M$ alone, because $Q=q^2\e^{2\Phi_0}/(2M)$ changes when mass and charge are radiated. The radiation channels should include neutral scalar modes, charged modes, and possibly dilaton perturbations. The leading expectation is that the temperature prescription controls the overall scale of the flux, while $Q$ controls both the area and the greybody barrier.

\section{Discussion and Outlook}\label{sec:discussion}

The Gao--Zhang dilaton--de Sitter solution sits at the intersection of two well-developed subjects: dilaton black holes and de Sitter horizon thermodynamics. The former literature has clarified the geometry, scalar hair, and several perturbative properties of the solution; the latter has shown that Hawking radiation in a static de Sitter patch is sensitive to the normalization of the timelike Killing vector and to the chosen effective two-horizon thermodynamic description. The comparison carried out here indicates that these two facts cannot be combined by a trivial replacement of the Schwarzschild areal radius. The dilaton modifies the entropy, the singularity condition, the photon sphere, and the greybody potential, while the de Sitter temperature ambiguity changes the Planck factor and the inferred evaporation rate.

Several calculations would turn this formulation into a complete quantitative study. First, one should compute the greybody factors $\Gamma_\ell(\omega)$ numerically for the potential in Equation~\eqref{eq:scalar_potential} across the allowed $(M,Q,\Lambda)$ domain. Second, the same calculation should be repeated for electromagnetic and dilaton perturbations, since the scalar hair makes the radiated-channel bookkeeping more subtle than in Schwarzschild--de Sitter. Third, the first law should be derived in a quasilocal ensemble that treats the two horizons, electric charge, scalar charge, and cosmological pressure consistently. Finally, the evaporation equations should be integrated with different temperature prescriptions to determine whether the endpoint is a Nariai configuration, a cold dilatonic remnant, or a naked singularity avoided by charge loss.

The central conclusion is that the thermodynamics of these black holes has not been exhausted by assigning $T=|f'(r_h)|/(4\pi)$. In de Sitter space that assignment is one prescription among several, and for dilaton black holes the differences are amplified by the scalar-dependent area and redshift structure. A dedicated study of the Gao--Zhang geometry along these lines would therefore fill a genuine gap in the literature.


\section*{Acknowledgments}
The author would like to thank B.~C.~L{\"u}tf{\"u}o{\u{g}}lu for fruitful scientific discussions and valuable comments.

		\section*{Funding}
This research received no external funding. 
 
		\section*{Institutional Review Board Statement}
Not applicable. 

		\section*{Informed Consent Statement}
Not applicable. 


		\section*{Data Availability Statement}
Not applicable. 

		\section*{Conflicts of Interest}
The authors declare no conflict of interest. 
 



\section*{Use of AI and AI-Assisted Technologies}
During the preparation of this manuscript, the author employed ChatGPT (GPT-5, OpenAI) to assist in the refinement of language and improvement of textual clarity and style. After using this tool/service, the author reviewed and edited the content as needed and takes full responsibility for the content of the published article.

	\small
	\bibliographystyle{scilight}
	
	

\end{document}